\documentclass[manuscript]{acmart}
\usepackage{booktabs} % For formal tables
\usepackage[font=small]{caption}
\usepackage[labelformat=simple]{subcaption}
\usepackage{graphicx}
\usepackage[ruled,vlined,linesnumbered]{algorithm2e}
\usepackage{algorithmic}
\SetKwComment{Comment}{$\triangleright$\ }{}
\usepackage[mathscr]{eucal}
\usepackage{color}
\usepackage{url}
\usepackage{mathtools}
\usepackage{amsfonts}
\usepackage{xcolor}
\usepackage{multirow}
\usepackage{soul}
\usepackage{colortbl}

\allowdisplaybreaks

\DeclareMathAlphabet{\mathbcal}{OMS}{cmsy}{b}{n}

\newcommand{\method}{\textsc{HierSRec}\xspace}
\newcommand{\tagnn}{\textsc{TAGNN}\xspace}
\newcommand{\mtrec}{\textsc{M2TRec}\xspace}
\newcommand{\cotrec}{\textsc{COTREC}\xspace}

\newcommand{\narm}{\textsc{NARM}\xspace}
\newcommand{\stamp}{\textsc{STAMP}\xspace}

\newcommand{\csrm}{\textsc{CSRM}\xspace}

\newcommand{\hide}[1]{}
\begin{document}

\settopmatter{printacmref=false}
\setcopyright{none}
\copyrightyear{2023}
\acmYear{2023}
\acmDOI{10.1145/1122445.1122456}
\acmConference[RecSys ORSUM Workshop '23]{The 6th Workshop on Online Recommender Systems and User Modeling at RecSys 2023}{September 18-22, 2023}{Singapore}
\acmPrice{15.00}
\acmISBN{978-1-4503-XXXX-X/18/06}

\title{Hierarchical Multi-Task Learning Framework  for Session-based Recommendations}


\author{Sejoon Oh}
\authornote{This work is based on the SO's internship project at The Home Depot.}
\email{soh337@gatech.edu}
\affiliation{%
	\institution{Georgia Institute of Technology}
	%%  \streetaddress{P.O. Box 1212}
	  \city{Atlanta}
      \state{Georgia}
	\country{USA}
	%%  \postcode{43017-6221}
}

\author{Walid Shalaby}
%% \authornote{Both authors contributed equally to this research.}
\email{walid_shalaby@homedepot.com}
%% \orcid{1234-5678-9012}
%% \author{Amir Afsharinejad}
%% \authornotemark[1]
%% \email{webmaster@marysville-ohio.com}
\affiliation{%
  \institution{The Home Depot}
%%  \streetaddress{P.O. Box 1212}
	  \city{Atlanta}
      \state{Georgia}
 \country{USA}
%%  \postcode{43017-6221}
}

\author{Amir Afsharinejad}
\email{amir_afsharinejad@homedepot.com}
\affiliation{%
  \institution{The Home Depot}
%%  \streetaddress{P.O. Box 1212}
	  \city{Atlanta}
      \state{Georgia}
 \country{USA}
%%  \postcode{43017-6221}
}

\author{Xiquan Cui}
\email{xiquan_cui@homedepot.com}
\affiliation{%
	\institution{The Home Depot}
	%%  \streetaddress{P.O. Box 1212}
	  \city{Atlanta}
      \state{Georgia}
	\country{USA}
	%%  \postcode{43017-6221}
}



\renewcommand{\shortauthors}{Oh et al.}


% and recommend relevant items according to the current interest of the user

\begin{abstract}
While session-based recommender systems (SBRSs) have shown superior recommendation performance, multi-task learning (MTL) has been adopted by SBRSs to enhance their prediction accuracy and generalizability further.
Hierarchical MTL (H-MTL) sets a hierarchical structure between prediction tasks and feeds outputs from auxiliary tasks to main tasks. This hierarchy leads to richer input features for main tasks and higher interpretability of predictions, compared to existing MTL frameworks. However, the H-MTL framework has not been investigated in SBRSs yet. 
In this paper, we propose \method which incorporates the H-MTL architecture into SBRSs. 
\method encodes a given session with a metadata-aware Transformer and performs next-category prediction (i.e., auxiliary task) with the session encoding. Next, \method conducts next-item prediction (i.e., main task) with the category prediction result and session encoding.
For scalable inference, \method creates a compact set of candidate items (e.g., 4\% of total items) per test example using the category prediction.
Experiments show that \method outperforms existing SBRSs as per next-item prediction accuracy on two session-based recommendation datasets. 
The accuracy of \method measured with the carefully-curated candidate items aligns with the accuracy of \method calculated with all items, which validates the usefulness of our candidate generation scheme via H-MTL.
% We also perform an ablation study and a case study of \method to verify the effectiveness and interpretability of our proposed H-MTL architecture for SBRSs, respectively.
\end{abstract}
\maketitle

\section{\textbf{Introduction}}
\label{sec:intro}

\noindent \textbf{Problem Description and Motivation.} 
Multi-task learning (MTL)~\cite{qiu2021incorporating,huang2021graph,meng2020incorporating,gao2019neural, hadash2018rank} has been employed to enhance the accuracy of existing recommender systems. MTL prevents the overfitting of a model via sharing parameters between multiple prediction tasks~\cite{ruder2017overview}. Hierarchical MTL (H-MTL)~\cite{sanh2019hierarchical,wang2021end,tian2019hierarchical,park2019hierarchical} further improves the MTL by exploiting predictions from other tasks as another task's input in a hierarchical order.  For example, the main task (e.g., next-item prediction) can use outputs from auxiliary tasks (e.g., next-category prediction) as input features to its prediction model. Those additional features can serve as rich external knowledge and enhance the performance of the main task.

While such an H-MTL framework gives an implicit data augmentation effect and higher generalization capability to a machine learning model~\cite{wang2021end, park2019hierarchical}, its application to session-based recommender systems (SBRSs)~\cite{li2017neural,liu2018stamp,wang2019collaborative,hansen2020contextual,wu2019session,hidasi2018recurrent, you2019hierarchical,pan2020intent,ren2019repeatnet} has not been investigated yet.
SBRSs have gained attention as they capture the latest and evolving interests of a user in a session, where a session consists of a sequence of user-item interactions occurring within a short period~\cite{wang2019survey}.
The H-MTL architecture will be beneficial and important to SBRSs as per not only prediction accuracy~\cite{lim2022hierarchical} but also interpretability~\cite{chen2019co}. Let us assume a user is searching for a product to add to the current session in an e-commerce platform. With the H-MTL, we not only enhance the recommendation quality to users by leveraging prior knowledge obtained from auxiliary tasks but also provide meaningful and reasonable explanations of the recommendations to users by interpreting outputs from auxiliary tasks (e.g., top-K predicted categories or interaction types).

\noindent \textbf{Challenges.}
Devising the H-MTL framework optimized for SBRSs is challenging for three major reasons. First, we need to define appropriate auxiliary tasks (e.g., category or interaction type predictions) related to the main task; in addition, we ought to set up proper hierarchical relationships between prediction tasks (e.g., a bottom-up approach from next-category prediction to next-item prediction). Second, rich and accurate session representations are required to ensure the high accuracy of the prediction tasks. The session features should contain item metadata information (e.g., categories) so that they can be used for auxiliary prediction tasks. Finally, it is computationally prohibitive to test the performance of H-MTL for SBRSs with millions of items available on online platforms (e.g., Amazon website). While existing methods~\cite{li2020time,kang2018self,sun2019bert4rec} use randomly-sampled candidate items for the test, the sampled metrics can be inconsistent with the original performance measured with all items~\cite{krichene2020sampled}. Thus, how can we generate high-quality candidate items to accurately evaluate the performance of H-MTL for SBRSs?

\begin{table}[t!]
    \footnotesize
	\centering
	\caption{Comparison of our proposed recommendation framework \method to existing session-based recommender systems.
	}
	\begin{tabular}{c|c|cccccc}
		\toprule
		& \textbf{\method} 
        &  \narm~\cite{li2017neural}  
		&   \stamp~\cite{liu2018stamp}
		&  \csrm~\cite{wang2019collaborative} &  \tagnn~\cite{yu2020tagnn} &  \cotrec~\cite{COTREC}  &
		\mtrec~\cite{shalaby2022m2trec}
		 \\
		\midrule
		Session-based Recommendation & \checkmark &  \checkmark & \checkmark & \checkmark &  \checkmark & \checkmark & \checkmark \\
		Employing Item Metadata & \checkmark &   &     &  & & & \checkmark      \\
		Multi-task Learning & \checkmark &   &     &  &   & \checkmark  & \checkmark   \\
		Hierarchical Learning & \checkmark &   &  & \checkmark  &   & &   \\
		Candidate Generation for Test & \checkmark  &   &  &  &  &  & \\
		\bottomrule
	\end{tabular}
	\label{tab:comparators}
\end{table}
\noindent \textbf{Proposed Method.}
To address the above challenges, we propose a novel recommendation model called \method which incorporates the H-MTL framework to SBRSs. 
\method is trained with multiple objectives of predicting the next item (main task) and next category (auxiliary task) in a session.
We choose next-category predictions as our auxiliary task since category labels are mostly available in recommendation datasets (e.g., Amazon product review~\cite{ni2019justifying}, Diginetica~\cite{Diginetica}). If categories labels are partially available or completely unavailable, we can cluster items to obtain implicit category information~\cite{oh2022implicit} or predict other metadata in a dataset such as interaction types (e.g., purchase, click, etc.) as auxiliary tasks. 
The first step of \method is generating a session representation using a metadata-aware Transformer~\cite{vaswani2017attention} encoder. The Transformer encoder uses item IDs, item categories, and additional item metadata such as titles and descriptions to produce a precise and rich summary of the session. After that, \method predicts the next category using the session encoding vector and transforms those category prediction results into embeddings. Next, we predict the next item in a session using the session representation and category prediction embeddings. 
Finally, the next-item and next-category prediction losses are combined together to train \method.
After training, we first perform the category prediction for all test instances, and we can generate high-quality candidate items for each test example by aggregating items belonging to top-K predicted categories.

\noindent \textbf{Experiments.}
Thorough experiments on two large-scale E-commerce datasets show that \method has superior next-item prediction performance to existing SBRSs by leveraging the hierarchical prediction framework.
\method shows at least $6.7\%$ performance improvements as per three accuracy metrics  compared to baselines. 
Moreover, \method achieves comparable accuracy to that of \method tested with full items by deliberately selecting a few items (e.g., $4\%$ of total items) as ranking candidates using the category prediction.
Ablation studies of \method verify the effectiveness of each component of \method.

\noindent \textbf{Contributions.} The main contributions of our paper are summarized as follows.

\begin{itemize}
    \item To the best of our knowledge, this is the first work to leverage the H-MTL framework for SBRSs. 
    \item We propose \method that accurately predicts the next item in a session via employing the output of next-category prediction. \method offers a compact set of candidate items of each test example for scalable ranking.
    \item Experiments on two recommendation datasets show that \method outperforms existing SBRSs as per next-item prediction accuracy. We also confirm the effectiveness of our candidate generation method.
\end{itemize}

% \vspace{-2mm}
\section{\textbf{Related Work}}
\label{sec:related_work}
\noindent \textbf{Multi-Task Learning (MTL) \& Hierarchical MTL.}
Recent research has shown that the generalizability and prediction performance of recommendation models can gain substantial improvements by using multi-task learning (MTL)~\cite{gao2019neural, hadash2018rank, caruana1997multitask}. In particular, MTL shares the knowledge learned from other related tasks with the main one, which has been shown to not only enhance the overall performance of the model, but also decrease the chance of overfitting and improve the quality of learned representations \cite{ruder2017overview}. 
Hierarchical learning can improve the generalizability and interpretability of MTL by using the predictions of related tasks for another task. This architecture is called Hierarchical MTL (H-MTL). H-MTL has been utilized in Natural Language Processing~\cite{sanh2019hierarchical, wang2021end,  tian2019hierarchical, farag2019multi, song2022enhance,nguyen2019multitask,song2020hierarchichal} and Computer Vision~\cite{park2019hierarchical,nguyen2019multitask, fan2017hd, yan2015hd} domains to boost the performance of a model by sharing knowledge from lower-level tasks for more complex ones. In the context of recommender systems, \citet{chen2019co} use H-MTL to improve the prediction accuracy and also provide a linguistic explanation of why a user likes/dislikes an item. \citet{lim2022hierarchical} also utilize H-MTL to predict the Point-of-Interest (POI) a user will visit next. 

 \noindent \textbf{Session-based Recommendation.}
Neural networks have served as the key component of the state-of-the-art SBRSs. Recurrent neural networks (RNNs) have been used to capture item dependencies within sessions~\cite{li2017neural,liu2018stamp, wang2019collaborative, pan2020intent, ren2019repeatnet}. However, RNNs are limited in capturing longer dependencies across items. Thus, graph neural network-based approaches~\cite{wu2019session, yu2020tagnn, COTREC,  xu2019graph, wang2020global} and attention-based methods~\cite{li2017neural,liu2018stamp, wang2019collaborative, pan2020intent, ren2019repeatnet} have been proposed to incorporate such dependencies precisely into SBRSs. Transformer~\cite{vaswani2017attention}-based approaches provide  superior performance in predictions~\cite{de2021transformers4rec, de2021end, chen2019bert4sessrec, moreira2021transformers}. 
MTL has been utilized to improve the accuracy of next-item prediction in SBRSs~\cite{tavakol2014factored, qiu2021incorporating, huang2021graph, meng2020incorporating,shalaby2022m2trec}. 
User intent prediction~\cite{guo2022learning, oh2022implicit, liu2021intent} is a well-known example of applying MTL to SBRSs.
However, existing SBRSs are either designed only for next-item prediction or incompatible with the H-MTL architecture.

\section{Proposed Approach: \method}
\label{sec:proposed_method}
\begin{figure*}[t!]
    \centering
    \includegraphics[width=1.0\linewidth]{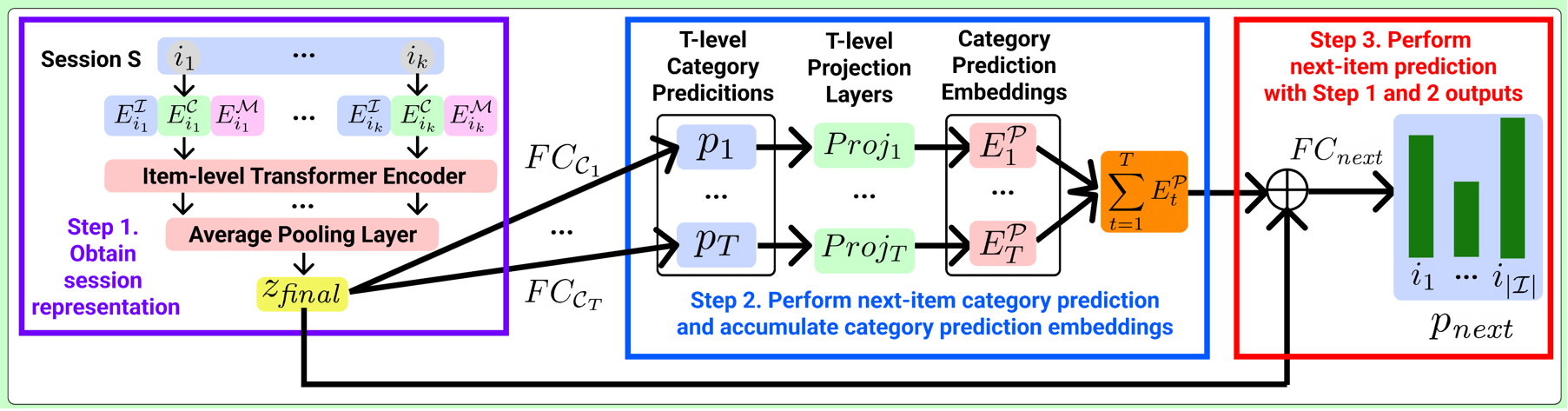}
    \caption{
    \textit{Overview of \method}. Given a session and its observed items, \method first generates a session representation via a Transformer encoder, and it performs next-category and next-item predictions in a hierarchical manner using the session representation. 
    }
    \label{fig:H-MTL}
\end{figure*}

\noindent \textbf{Overview.}
As shown in Figure~\ref{fig:H-MTL}, our proposed recommendation model \method employs a Transformer~\cite{vaswani2017attention} architecture to encode items in a session accurately and utilizes the obtained session representation for the \textit{\textbf{hierarchical 
MTL}}~\cite{sanh2019hierarchical, wang2021end, tian2019hierarchical, park2019hierarchical, fan2017hd}. We define tasks of \method as the next item prediction (main) and multi-level category predictions (auxiliary), where the category information is available on many recommendation datasets (e.g., Amazon product review~\cite{ni2019justifying}, Diginetica~\cite{Diginetica}). Notice that our framework can be easily extended to other types of auxiliary tasks such as predicting user actions (e.g., click, add-to-cart, or purchase).
Finally, we offer a candidate item generation scheme that uses the output from auxiliary tasks for scalable inference or evaluation.

\noindent \textbf{Session Encoding with Metadata-aware Transformer.}
We use the Transformer encoder~\cite{vaswani2017attention} to create an accurate representation of a user's interest within a session. Formally, given a session $\mathcal{S}$ with a sequence of $k$ observed items $\{i_1, \ldots, i_{k}\}$, the first step is transforming the item sequence  to the item representation sequence $\{E_{i_1}, \ldots, E_{i_{k}}\}$ using the item ID, the item category information, and the other item metadata such as titles and descriptions. Given an observed item $i_{pos}, \forall pos, 1 \leq pos \leq k$ in the current session $\mathcal{S}$, its embedding is constructed as follows.
\begin{equation}
\centering
   E_{i_{pos}} = TransEnc(concat(E_{i_{pos}}^{\mathcal{I}}, E_{i_{pos}}^{\mathit{\mathcal{C}}}, E_{i_{pos}}^{\mathcal{M}})),
   \label{eq:item_embedding}
\end{equation}
where $TransEnc$ and $concat$ indicate the Transformer encoder and embedding concatenation operation, and $E^{\mathcal{I}}$, $E^{\mathcal{C}}$, and $E^{\mathcal{M}}$ denote item ID embeddings, item category embeddings, and item metadata embeddings, respectively.
Different encoders such as STAMP~\cite{liu2018stamp} can be used as $TransEnc$ in \method, but our metadata-aware Trasnformer shows the best empirical performance.
Given a set of items $\mathcal{I}$ and item ID embedding dimension $d_{\mathcal{I}}$, item ID embeddings $E^{\mathcal{I}} \in \mathbb{R}^{|\mathcal{I}| \times d_{\mathcal{I}}}$ map each item ID to the $d_{\mathcal{I}}$-dimensional feature.
Assuming there are $T$-level item categories (e.g., 3-level categories for a \textit{``dinner plate''} item are $\mathcal{C}_1$ - Kitchen; $\mathcal{C}_2$ - Tableware \& Bar; $\mathcal{C}_3$ - Dinnerware), we create $d_{\mathcal{C}}$-dimensional embeddings for all category levels (i.e.,  $E^{\mathcal{C}_1}, \ldots, E^{\mathcal{C}_T}$)  and sum up all category embeddings of an item to encode its category information, i.e., $E_{i_{pos}}^{\mathcal{C}} = E_{i_{pos}}^{\mathcal{C}_1} + \ldots + E_{i_{pos}}^{\mathcal{C}_T}$ .
Finally, item metadata embeddings $E^{\mathcal{M}} \in \mathbb{R}^{|\mathcal{I}| \times d_{\mathcal{M}}}$ are concatenations of trainable or pre-trained embeddings (e.g., from BERT~\cite{devlin-etal-2019-bert} for texts and ResNet~\cite{he2016deep} for images) for each metadata of an item. For instance, the metadata embedding of an item $i_{pos}$ can be derived as follows: $E^{\mathcal{M}}_{i_{pos}} = concat(E^{\text{title}}_{i_{pos}},E^{\text{description}}_{i_{pos}},E^{\text{image}}_{i_{pos}})$.

Finally, the item representation sequence $\{E_{i_1}, \ldots, E_{i_k}\}$ generated by the Transformer is fed to the average pooling layer to generate an accurate session representation $z_{\textit{final}}$ of a given session $\mathcal{S}$ (i.e., $z_{\textit{final}} \in \mathbb{R}^{d_{\mathcal{I}}+d_{\mathcal{C}}+d_{\mathcal{M}}} = \mathit{pooling}(E_{i_1}, \ldots, E_{i_k})$).
We choose the average pooling since it empirically shows the best next-item prediction performance compared to the trainable pooling or max pooling.

\noindent \textbf{Hierarchical Multi-Task Learning (H-MTL).}
Given the final session representation $z_{\textit{final}}$ of a session $\mathcal{S} = \{i_1, \ldots, i_{k}\}$, a simple yet effective way to predict the next item $i_{k+1}$ in $\mathcal{S}$ is employing a fully-connected layer to transform $z_{\textit{final}}$ to a next-item prediction score vector.  However, this approach can easily make a model overfit the training data compared to multi-task learning (MTL) with sharing representations~\cite{ruder2017overview, chowdhuri2019multinet,wang2009boosted, zhang2019deep}.

While the MTL method can avoid the overfitting problem, it can be further enhanced by introducing a \textit{``hierarchy''} or order between the prediction tasks, which is called hierarchical MTL (H-MTL)~\cite{sanh2019hierarchical, wang2021end, tian2019hierarchical, park2019hierarchical, fan2017hd}.  
H-MTL models fully exploit the outputs from other tasks as \textit{``additional knowledge''} via performing predictions of multiple tasks in a specific order or hierarchically (i.e., tree-structure), which offers the implicit data augmentation effect and the generalization capability to the main model.

To adapt H-MTL to SBRSs,
we first predict categories of the next item (e.g., $i_{k+1}$) in a session  $\mathcal{S} = \{i_1, \ldots, i_{k}\}$ and employ the category prediction results to enhance the item prediction. Specifically, we obtain $T$-level category prediction score vectors (i.e., $\{p_{1}, \ldots, p_{T}\}$) of the next item in a session using a session representation $z_{\textit{final}}$, as shown below.
\begin{equation}
\centering
    p_{t} = FC_{\mathcal{C}_t}(z_{\textit{final}}) \in \mathbb{R}^{|\mathcal{C}_t|}, \forall t,  1 \leq t \leq T
    \label{eq:category_prediction},
\end{equation}
where $FC$ indicates a fully-connected layer.
Next, we transform category prediction vectors $\{p_{1}, \ldots, p_{T}\}$ to category prediction embeddings $\{E^{\mathcal{P}}_{1}, \ldots, E^{\mathcal{P}}_{T}\}$ via projection layers: $\mathit{Proj}_{t} \in \mathbb{R}^{|\mathcal{C}_t| \times (d_{\mathcal{I}}+d_{\mathcal{C}}+d_{\mathcal{M}})}, \forall t, 1 \leq t \leq T$, as shown below.
\begin{equation}
\centering
    E^{\mathcal{P}}_{t} = \mathit{Proj}_{t}(p_{t}) \in \mathbb{R}^{(d_{\mathcal{I}}+d_{\mathcal{C}}+d_{\mathcal{M}})}, \forall t,  1 \leq t \leq T.
    \label{eq:category_projection}
\end{equation}
Finally, we sum the session representation $z_{\textit{final}}$ and all category prediction embeddings $\{E^{\mathcal{P}}_{1}, \ldots, E^{\mathcal{P}}_{T}\}$ and feed it to the fully-connected layer to generate the next-item prediction vector $p_{\mathit{next}} \in \mathbb{R}^{|\mathcal{I}|}$, as shown below. 
\begin{equation}
\centering
p_{\mathit{next}} = FC_{\mathit{next}}(z_{\textit{final}}+\lambda_{C} \sum_{t=1}^{T}{E^{\mathcal{P}}_{t}}),
    \label{eq:next_item_prediction}
\end{equation}
where $\lambda_{C}$ (a hyperparameter) controls the impact of category predictions on the next-item prediction.
Category prediction results $\{p_{1}, \ldots, p_{T}\}$ can be used as implicit explanations of the next-item prediction $p_{\mathit{next}}$ as the category prediction embeddings are used as a part of input features to the next-item predictor.

\noindent \textbf{Loss Function and Optimization.}
Since \method is a multi-task learning method, loss functions of multi-level next-category prediction and next-item prediction are combined and jointly optimized together.
We use the Cross-Entropy loss for both category and item predictions.
Assuming the ground-truth $T$-level categories of a next-item $i_{k+1}$ we want to predict are $c_{1}, \ldots, c_{T}$, then a loss function of a level-$t$ category prediction task is given as follows.
\begin{equation}
    \begin{aligned}
        \mathcal{L}_{\mathcal{C}_{t}}(p_{t}) = -\sum_{i=1}^{|\mathcal{C}_t|} {y_{t}(i) \log \bigg(\mathit{Softmax}(p_{t})_{i}\bigg)},
    \end{aligned}
    \label{eq:loss_next_category}
\end{equation}
where $y_{t}$ is a one-hot vector whose $c_{t}^{th}$ value is 1, $p_{t}$ is a level-$t$ category prediction score vector, and $\mathit{Softmax}(x)_{i} = \frac{e^{x_{i}}}{\sum_{j}{e^{x_{j}}}}$.
Similarly, the next-item prediction loss is given as follows.
\begin{equation}
    \begin{aligned}
         \mathcal{L}_{\mathit{next}}(p_{\mathit{next}}) = -\sum_{i=1}^{|\mathcal{I}|} {y(i) \log \bigg(\mathit{Softmax}(p_{\mathit{next}})_{i}\bigg)},
    \end{aligned}
    \label{eq:loss_next_item}
\end{equation}
where $y$ is a one-hot vector whose $i_{k+1}^{th}$ value is 1, and $p_{\mathit{next}}$ is a next-item prediction score vector.
The combined loss function for our proposed hierarchical multi-task learning is given as follows.
\begin{equation}
    \begin{aligned}
         \mathcal{L}_{\mathit{final}}=  \mathcal{L}_{\mathit{next}}+\lambda \sum_{t=1}^{T}  \mathcal{L}_{\mathcal{C}_{t}},
    \end{aligned}
    \label{eq:loss_MTL}
\end{equation}
where $\lambda$ is the importance weight of category prediction tasks. 
We tested several weighting strategies for $\lambda$ and $\lambda_{C}$ such as trainable weights or randomized weights per epoch, but \textbf{setting $\lambda =  \lambda_{C} = 1.0$ shows the best prediction performance empirically}.
We optimize the above loss function~\eqref{eq:loss_MTL} with Adam~\cite{kingma:adam} optimizer for all training data. 

\noindent \textbf{Candidate Item Generation for Scalable Evaluation.}
During the test (or inference) stage, we use the next-item prediction vector $p_{\mathit{next}} \in \mathbb{R}^{|\mathcal{I}|}$ generated from the trained model. 
Computing $p_{\mathit{next}}$ can be computationally expensive on large-scale recommendation datasets (e.g., e-commerce domain) with millions of items. Thus, for practicality, existing algorithms~\cite{li2020time,kang2018self,sun2019bert4rec} uses a small set of candidate items instead of all items during the inference. However, those methods use randomly-sampled candidates, and the accuracy measured with such candidates can be significantly different from the accuracy calculated with full items~\cite{krichene2020sampled}.
Thus, we propose a more accurate candidate generation method that leverages the category prediction result. 
First, given a test session with observed items, we conduct the category prediction and obtain the score vectors $\{p_{1}, \ldots, p_{T}\}$. We sample top-K (K: hyperparameter) categories $\{c_{1}^{t}, \ldots, c_{K}^{t}\}$ from each level-$t$ category prediction $p_{t}$ and construct a candidate item set $\mathcal{I'} \subset \mathcal{I}$ by the following.
$$
\mathcal{I'} = \{ i \mid i \in c_{k}^{1}, \forall k, 1 \leq k \leq K\} \cup \cdots \cup  \{ i \mid i \in c_{k}^{T}, \forall k, 1 \leq k \leq K\}
$$
We empirically verify that our candidate selection policy can achieve nearly equivalent accuracy to that of an original policy that uses all items during the inference (refer to Figure~\ref{fig:case_study} later).

\noindent \textbf{What If Category Information Is Unavailable?}
While item category information is available in most recommendation datasets, it might be partially available or completely unavailable in a few cases. To address it, we can find implicit categories of items by utilizing graph neural networks and clustering~\cite{oh2022implicit}. For instance, we apply an off-the-shelf node embedding algorithm on a session-item bipartite graph to obtain item representations. Applying a clustering method (e.g., K-means) on obtained item embeddings will generate clusters of items, which will approximate the category information. 
Another solution is employing other metadata (e.g., interaction type) for auxiliary tasks of the H-MTL.

\section{Experimental Evaluations of \method}
\label{sec:exp}
	\begin{table}[t!]
 \small
		\caption{Summary of datasets and sessions used for experiments. M: million, K: thousand. N/A: hidden due to the company's policy. }
		\begin{tabular}{| c | c | c | c | c | c | c | c|}
			\toprule
			\textbf{Name} & \textbf{\begin{tabular}[c]{@{}c@{}}Training \\ sessions\end{tabular}} & \textbf{\begin{tabular}[c]{@{}c@{}}Test \\ sessions\end{tabular}} & \textbf{Items}  &
   \textbf{Interactions}  &
   \textbf{Session Length}  &
      \textbf{Category}  &
   \textbf{\begin{tabular}[c]{@{}c@{}}Item \\ Metadata\end{tabular}}  
            \\
			\midrule
		\begin{tabular}[c]{@{}c@{}}The Home Depot (THD) \\ (E-commerce)\end{tabular}  & 2.8M & 85K  & 259K & N/A & N/A & N/A & \begin{tabular}[c]{@{}c@{}}Title, Description, \\ Category, etc.\end{tabular}\\
			\hline 
			\begin{tabular}[c]{@{}c@{}}Diginetica \\ (E-commerce)\end{tabular} & 191K & 16K & 117K  & 880K & 4.60 & Level-1: 1.2K & Title, Category\\
		\bottomrule
		\end{tabular}	
		\label{tab:dataset}
	\end{table}
 
\begin{table*}[t!]
	\centering
    \large
	\caption{Performance of \method in terms of predicting the next item in a session on the Diginetica and THD datasets, compared to baseline SBRSs (\textbf{Bold} indicates the best model, while the second-best model is \underline{underlined}). \method shows the best prediction performance among all methods across two datasets, with statistical significance (P-values from one-tailed t-test are $\leq 0.05$).}
	\begin{tabular}{|l|c|c|c|c|c|c|}%c|c|c}
	\hline
	Dataset  & \multicolumn{3}{c|}{Diginetica} & \multicolumn{3}{c|}{The Home Depot} \\%& \multicolumn{3}{c}{Yoochoose} \\
	\hline
	Method  & MRR@20 & HITS@20 & Recall@20 & MRR@20 & HITS@20 & Recall@20 \\%& HIT@20 & R@20 & MRR@20 \\
	\hline
	\narm~\cite{li2017neural} & 0.0751 & \underline{0.2993} & \underline{0.4621} & 0.1036 & \underline{0.2792} & \underline{0.4163} \\%& & & \\
	\stamp~\cite{liu2018stamp} & 0.0717 & 0.2545 & 0.4161 & \underline{0.1086} & 0.2718 & 0.3996 \\%& & & \\
	\csrm~\cite{wang2019collaborative} & 0.0730 & 0.2804 & 0.4363 & 0.0968 & 0.2608 & 0.3919\\%& & & \\
	\tagnn~\cite{yu2020tagnn} & 0.0785 & 0.2643 & 0.4153 & 0.0865 & 0.2188 & 0.3507\\%& & & \\
	\cotrec~\cite{COTREC} & \underline{0.0787} & 0.2837 & 0.4306 & 0.0625 & 0.2271 & 0.3518\\%& & & \\
	%The Home Depot-Production & --- & --- & --- & 0.1072 & \underline{0.2834} & \underline{0.4214}  \\%& & & \\
	\hline
	\textbf{\method} & \textbf{0.0990} & \textbf{0.3347} & \textbf{0.5031} & \textbf{0.1245} & \textbf{0.3019} & \textbf{0.4443} \\%& & & \\
	%\% Improvement & 18.5\% & 7.4\% & 6.1\% & 14.6\%  & 6.5\% &  5.4\% \\
	\% Improvement & 25.8\% & 11.8\% & 8.9\% & 14.6\%  & 8.1\% &  6.7\% \\
	\hline
\end{tabular}
\label{tab:next_item_performance}
\end{table*}

\textbf{Datasets.} Table~\ref{tab:dataset} lists the statistics of the datasets. 
The Home Depot (THD) is an E-commerce dataset obtained from a large online retailer THD. The dataset is composed of Add-to-Cart (ATC) events within millions of online sessions. The dataset has rich product metadata including 7 attributes: product title, 3-level categories, brand, manufacturer, color, department name, and class name. We exclude certain information from the THD dataset according to the company's policy.
For the THD dataset, we filter users and items with less than 10 interactions.
Diginetica\footnote{https://competitions.codalab.org/competitions/11161} is a \textbf{public E-commerce dataset} that was a part of CIKM Cup 2016 challenge. We did pre-process of the Diginetica  similar to  \cite{wu2019session}.
 
\noindent
\textbf{Baselines.} We use the following state-of-the-art session-based recommenders:\footnote{We used open-source implementations of baseline algorithms (\url{https://github.com/rn5l/session-rec}).}  (1)  \textbf{\narm}~\cite{li2017neural}: An attention-based model that employs a hybrid encoder to reflect a user's global and local interests with an attention mechanism, (2) \textbf{\stamp}~\cite{liu2018stamp}: An attention/memory-based model that incorporates a user's short-term and long-term interests via short-term attention and long-term memory modules, respectively, (3) \textbf{\csrm}~\cite{wang2019collaborative}: a session-based recommendation model that contextualizes the current and neighborhood sessions with inner and outer memory encoders, respectively, (4) \textbf{\tagnn}~\cite{yu2020tagnn}: a graph neural network (GNN)-based session-based recommender that utilizes a target-aware attention module for predictions, and (5) \textbf{\cotrec}~\cite{COTREC}: a state-of-the-art GNN-based recommendation model that combines self-supervised learning with graph co-training. We exclude several models including nearest-neighbor algorithms if they show similar or worse performance compared to our existing baselines.

\noindent
\textbf{Hyperparameters.} 
Hyperparameters of \method and baseline methods are found by \textbf{extensive grid search} using a validation set (randomly sampled 10\% from training). Specifically, $d_{\mathcal{I}} = d_{\mathcal{C}} = d_{\mathcal{M}}=128$, $\lambda=\lambda_{C}=1.0$, and batch size and learning rates are set to $1024$ and $0.0001$, respectively.
We use all items as candidates during the test by default.
\method also uses 2 layers of Transformer Encoder with 8 attention heads. 

\noindent
\textbf{Reproducibility.}
While the code of \method and the THD dataset cannot be released due to the company policy, we release the public dataset (Diginetica) and baseline implementations used in the paper.

\noindent \textbf{Next-item Prediction Accuracy of \method.}
To verify the effectiveness of \method, we measure the next-item prediction accuracy of \method and baselines on diverse datasets, with respect to three accuracy metrics:  Mean Reciprocal Rank@20 (MRR)~\cite{voorhees1999trec}, HITS@20, and Recall@20. HITS@20 counts only ground-truth next-items in top-20 lists, while Recall@20 counts all future items (including the next-item) in a session in top-20 lists.

Table~\ref{tab:next_item_performance} shows the next-item prediction accuracy of \method and baselines on the Diginetica and THD datasets. \method shows the best performance as per all metrics among all methods across all datasets, with statistical significance (P-values from one-tailed t-test are $\leq 0.05$). Relative performance improvements of \method compared to the best baseline are $6.7\%-25.8\%$.
The high performance of \method is due to \textbf{the hierarchical learning architecture, not additional item metadata} (compare the first and last row in Table~\ref{tab:ablation_study}).
In other words, the key reason for these performance improvements is incorporating prior knowledge from the next-category prediction into the next-item prediction, so that we can filter out items associated with irrelevant categories easily while predicting the next item. 
\begin{table}[t!]
\small
\centering
\caption{\textit{Category prediction result (left) and ablation study (right) of \method on the Diginetica dataset.} Both results substantiate the usefulness of the proposed H-MTL framework used in \method for session-based recommendations.
}
\begin{subtable}{0.49\linewidth}
\centering
\caption{Category prediction result of \method.}
\begin{tabular}{|c|c|c|c|}
\hline
\textbf{Models / Metrics}  & \textbf{\begin{tabular}[c]{@{}c@{}}MRR \\ @20 \end{tabular}} & \textbf{\begin{tabular}[c]{@{}c@{}}HITS \\ @20 \end{tabular}} & \textbf{\begin{tabular}[c]{@{}c@{}}Recall \\ @20 \end{tabular}}\\ \hline
\begin{tabular}[c]{@{}c@{}}Heuristic  \\ (recommends \\ historical categories) \end{tabular}        &    0.7861     & 0.8793          &   0.8908        \\\hline
\begin{tabular}[c]{@{}c@{}}Multi-task learning \\ (no hierarchical \\ predictions) \end{tabular} 
&   \textbf{0.8803}      & \textbf{0.9338}          &     \textbf{0.9445}         \\ \hline
\textbf{\begin{tabular}[c]{@{}c@{}}\method \\ (proposed)  \end{tabular}}                                                   &  \underline{0.8747}     & \underline{0.9291}         &  \underline{0.9396}                     \\\hline
\end{tabular}
\label{tab:category_prediction}
\end{subtable}
\hspace{3mm}
\begin{subtable}{0.48\linewidth}
\centering
\caption{Ablation study of \method.}
\begin{tabular}{|c|c|c|c|}
\hline
\textbf{Models / Metrics}  &  \textbf{\begin{tabular}[c]{@{}c@{}}MRR \\ @20 \end{tabular}} & \textbf{\begin{tabular}[c]{@{}c@{}}HITS \\ @20 \end{tabular}} & \textbf{\begin{tabular}[c]{@{}c@{}}Recall \\ @20 \end{tabular}}\\ \hline
\begin{tabular}[c]{@{}c@{}}Single-task learning \\ (only next-item \\ predictions) \end{tabular}        &    0.0752    & 0.2524          &   0.4037       \\\hline
\begin{tabular}[c]{@{}c@{}}Multi-task learning \\ (no hierarchical \\ predictions) \end{tabular} 
&   \underline{0.0842}      & \underline{0.2991}          &     \underline{0.4689}         \\ \hline
\textbf{\begin{tabular}[c]{@{}c@{}}\method \\ (proposed)  \end{tabular}}                                                   &  \textbf{0.0990} & \textbf{0.3347} & \textbf{0.5031}             \\ \hline
\end{tabular}
\label{tab:ablation_study}
\end{subtable}
\end{table}

\noindent \textbf{Next-category Prediction Accuracy of \method.}
We test how accurately \method can predict the next category in a session on the Diginetica dataset. As shown in Table~\ref{tab:category_prediction}, \method outperforms the heuristic and shows almost the same accuracy as an MTL variant of \method without hierarchical learning. It is expected since the category prediction does not take any additional feature from the next-item prediction task due to its lower hierarchy, and enhancing the next-category prediction accuracy is not the main goal of \method.

\noindent \textbf{Ablation Study of \method.}
We conduct the ablation study of \method to show how effective the hierarchical learning of \method is for session-based recommendations. We create two variants of \method, where the first one only performs next-item predictions \textbf{only with metadata-aware Transformer (no MTL)}, and the second one employs normal MTL architecture without hierarchical predictions. 
Table~\ref{tab:ablation_study} shows the ablation study result of \method on the THD dataset. As we can notice, \method exhibits the highest accuracy compared to the two variants, with at least $7.3\%$ relative performance improvements and statistical significance. This result shows that our proposed hierarchical MTL architecture induces a higher generalization capability of a model and an implicit data augmentation effect. 
% We also confirm that a variant with normal MTL outperforms another variant without MTL, which shows the benefits of using MTL for SBRSs.

\begin{figure}[t!]
    \centering
     \begin{subfigure}[b]{0.49\linewidth}
    \includegraphics[width=7.5cm]{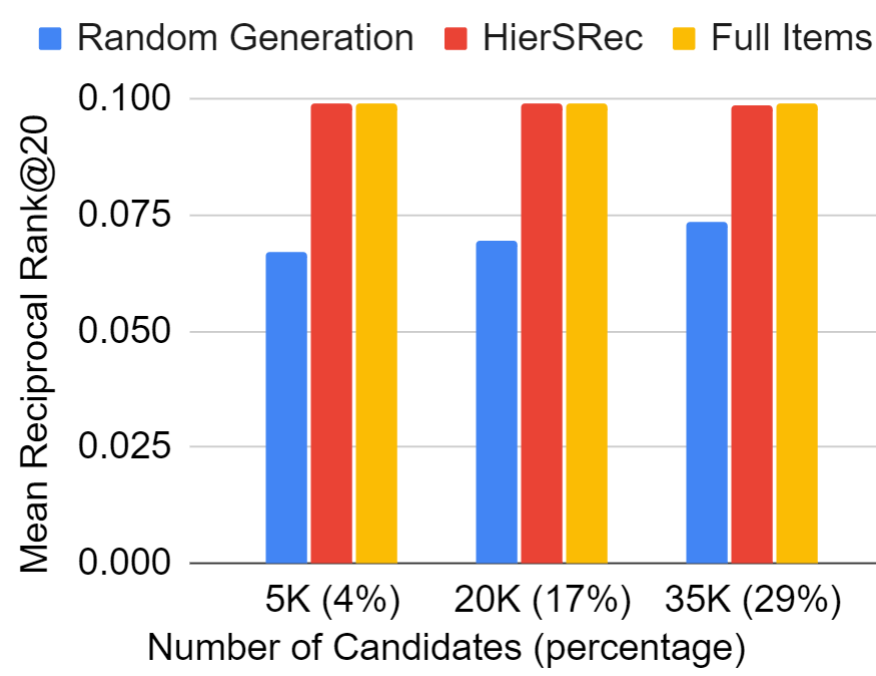}
 \caption{Next-item prediction accuracy of \method as per the number of candidates used for test.}
 \label{exp:candidate}
     \end{subfigure}
     \hfill
     \begin{subfigure}[b]{0.49\linewidth}
    \includegraphics[width=8cm]{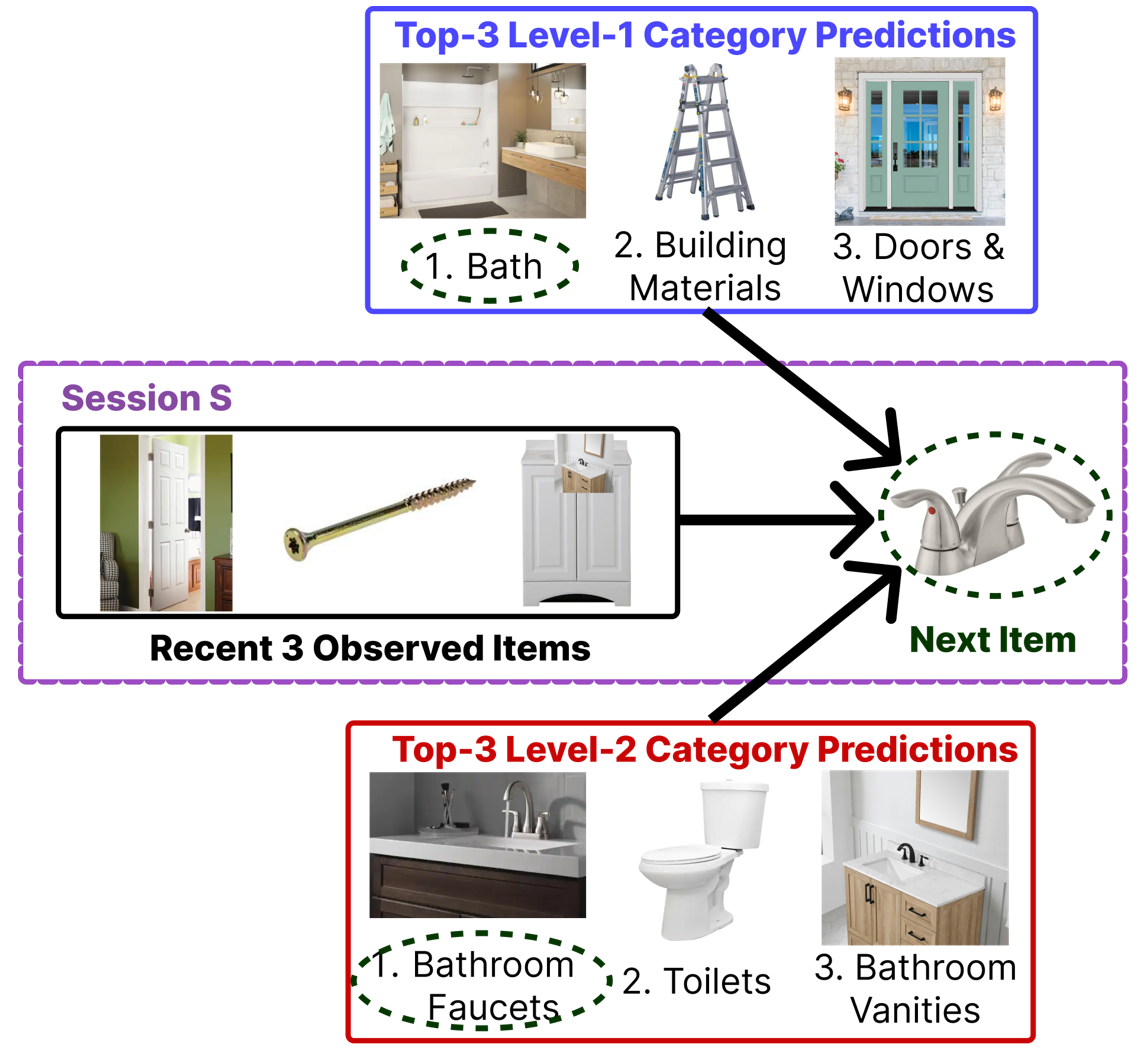}
 \caption{Case-study result of \method on the The Home Depot dataset.}
 \label{exp:interpretability}
     \end{subfigure} 
     \caption{\textit{Candidate generation verification (left) and case-study (right) results of \method.} 
    The left figure shows that accuracy measured with candidates generated by \method  is close to accuracy measured with full items in a dataset.
    The right figure indicates that the output of auxiliary tasks (e.g., Bath and Bathroom Faucets categories) can be crucial input features to the next-item prediction.
    }
	\label{fig:case_study}
\end{figure}

\noindent \textbf{Verification of Candidate Items Generated by \method.}
We confirm the quality of candidate items generated by \method by comparing the accuracy measured with our candidates, random candidates, and full items. Figure~\ref{exp:candidate} shows the MRR@20 metric of \method calculated with three different candidate generation policies on the Diginetica dataset. Using the category prediction knowledge, \method can create a small candidate set (e.g., 4\% of total items) consisting of key items that are highly related to the ground-truth next item in a session. However, the random candidate policy exhibits poor performance as it cannot selectively choose important items as well as the ground-truth next item.

\noindent \textbf{Case Study: Hierarchical Predictions of \method.}
Figure~\ref{exp:interpretability} is a case study result on the THD dataset of how \method utilizes category predictions to improve the next-item predictions. Top-K category prediction results of \method include diverse and evolving preferences of a user in a session (e.g., Door $\longrightarrow$ Bathroom) and identify the most important categories (e.g., Bath and Bathroom Faucets) for next-item predictions. 
Providing these meaningful knowledge from category predictions will make next-item recommendations personalized and accurate.

\section{Conclusion}\label{sec:conc}
In this paper, we proposed a novel session-based recommendation model \method that employs a metadata-aware Transformer encoder and a hierarchical multi-task learning framework to obtain higher model generalizability.
Future works of \method include that (1) more complex hierarchical structures (e.g., tree-shape) between various tasks (e.g., next-item, next-action, and next-category predictions) can be explored, and (2) extending \method to predict next items accurately in sessions with cold-start items or only a few items by employing their metadata. 
	
\bibliographystyle{ACM-Reference-Format}
\bibliography{references}

\end{document}